# *Sensor-less adaptive optics for Brillouin micro-spectroscopy*

Eitan Edrei and Giuliano Scarcelli

*Fischell Department of Bioengineering, University of Maryland, College Park, Maryland 20742, USA*

## Abstract

Brillouin spectroscopy is a powerful optical technique for viscoelastic characterization of samples without contact. However, like all optical systems, Brillouin spectroscopy performances are degraded by optical aberrations, and have therefore been limited to homogenous transparent samples. To correct for aberrations, adaptive optics (AO) methods have been previously integrated into a variety of optical modalities ranging from ground-based telescopes to super-resolution microscopes. In this work, we developed an adaptive optics configuration designed for Brillouin scattering spectral analysis. Our configuration doesn't require direct wave-front sensing and the injection of a 'guide-star'; hence, it can be implemented without the need for sample pre-treatment. We used our AO-Brillouin spectrometer in aberrated phantoms and biological samples; consistent with previous AO systems, we demonstrate effective correction of optical aberrations yielding 2.5-fold enhancement in Brillouin signal strength and 1.4-fold improvement in axial resolution.

## Introduction

For many decades, Brillouin light scattering spectroscopy, based on acoustic phonon-photon interaction [1], has been a powerful optical technique in applied physics and material science due to its unique ability to characterize mechanical properties of materials at high spatial resolution without contact [2-6]. More recently, Brillouin spectroscopy combined with confocal microscopy has found biological applications in cell biomechanics [7-13], plaque characterization [14] and is being tested in the clinic for ophthalmology applications [15, 16]. For Brillouin spectral measurements, two specifications are critical: due to the small spectral shift of Brillouin signatures, high spectral contrast (or extinction) is needed within spectrometers to eliminate noise from the incident laser or stray light; in addition, due to the small scattering cross-section of Brillouin interaction, the number of photons available for detection are fundamentally limited. Historically, spectrometers based on a cascade of Fabry-Perot etalons have provided sufficient extinction and resolution to detect and resolve Brillouin peaks [2, 11, 12]; yet, measurements with multi-pass Fabry-Perot interferometers are time consuming and generally not practical for imaging and/or biomedical applications. More recently, spectrometers based on virtually-imaged-phase-array (VIPA) etalon were developed to enable rapid measurements of Brillouin spectra [17, 18]. A vast amount of effort was dedicated in the past years to increase the spectral extinction of VIPA-based spectrometers; this progress has led to measurements of Brillouin spectra in non-transparent materials with shot-noise limited performances [19-23]. Thus, the next frontier of Brillouin microscopy progress is to improve signal to noise ratio (SNR) of Brillouin spectral measurements. In this context, multiplexed detection configurations [24] or improved spectral processing methods [25] have been explored as well as signal enhancement by stimulated Brillouin scattering process [26, 27]. All these methods, though, do not address the degradation of Brillouin signal due to optical aberrations. Aberrations are significant and unavoidable within optical systems, due to optical elements, sample inhomogeneity and refractive index mismatches [28]; hence, they impose a fundamental limitation on current capabilities of Brillouin spectrometers.

To address this limitation, we have developed a confocal Brillouin micro-spectrometer integrated with an adaptive optics (AO) system. AO is designed to measure and correct optical aberrations [29, 30] and has had great success in astronomy and ocular imaging providing aberration-free images of exosolar scenes [31] and retinal photoreceptors [32, 33]. Over the past twenty years, the progress of AO techniques has pushed imaging capabilities towards their fundamental limit, by enhancing SNR and providing higher resolution and contrast [34]. These advances have been traditionally focused to imaging modalities such as optical coherence tomography [35, 36], wide-field fluorescence [37-40], confocal [41-44] and multiphoton microscopy [43, 45-50].

In this work, we present an AO-Brillouin confocal system, designed to enhance the signal and resolution of Brillouin-based elasticity mapping through the correction of aberrations. The implementation of AO for Brillouin spectroscopy to improve both precision and resolution of elasticity mapping can potentially expand the application targets for Brillouin biomechanical studies.

## Principle

AO techniques can be classified in direct and indirect methods. Typically, direct methods use a wave-front sensor to measure the aberration while indirect methods either apply an iterative process to estimate the aberration or calculate the aberration from an acquired image. Direct approaches are faster because a single acquisition is needed to measure the phase aberration; however, the wave-front measurement requires a guide-star at the measured location, which may be difficult to introduce, especially in vivo [30]. In this work, we adopted an iterative indirect AO approach. Our optimization process is based on the acquired Brillouin signal and therefore doesn't require a 'guide star'. We used a confocal microscope system connected to a Brillouin spectrometer through an optical fiber working as a confocal pinhole. We placed a Spatial Light Modulator (SLM) within the illumination path of the confocal system; to enhance the Brillouin signal, we monitored the Brillouin spectrum intensity while varying the phase introduced by the SLM. The enhancement mechanism can be understood by considering the confocal configuration as shown in figure 1(a). Without aberrations, most of the Brillouin photons are generated at the illumination focal point which is conjugated to the confocal pinole (fig 1(a), first row), and therefore are transmitted to the spectrometer. However, an aberration within the optical path will disperse the incident light energy over a larger volume, and generate the Brillouin photons at various locations which are blocked by the confocal pinhole leading to a degraded signal (fig 1(a), second row). By correcting the aberration using a SLM, the focal point can be retrieved (fig 1(a), third row). Although aberration correction improves by optimizing also the collection path leading to the confocal pinhole, here we modified only the illuminating beam to improve overall throughput and minimize optimization time (see Discussion).

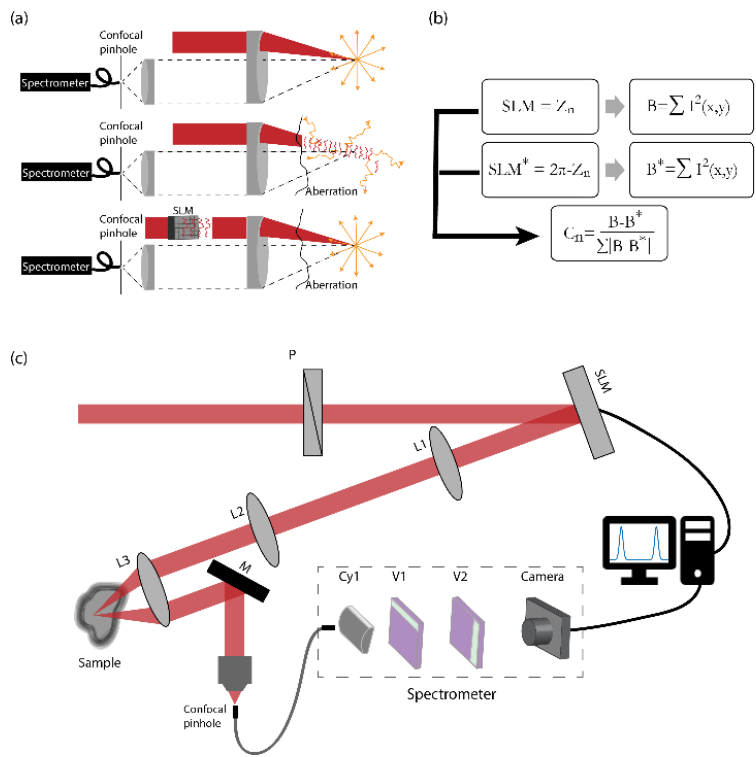

**Fig. 1**. AO optimization process and setup: (a) Enhancement mechanism: without aberrations the confocal pinhole collects all Brillouin photons originating from the conjugated focal point within the solid angle of the collecting lens (first row). As an aberration is introduced, the region in which the Brillouin photons are generated is extended, and therefore many Brillouin photons are blocked by the confocal pinhole (second row). By correcting the aberration, a sharp focal point is restored, and the signal is enhanced (third row). (b) Optimization steps: every Zernike polynomial and its inverse are projected on the SLM; the intensity of the spectrum is measured for both. The relative difference between the intensity values determines the magnitude of the measured Zernike polynomial. (c) Optical setup: A polarized expanded single frequency laser beam ($\lambda = 660nm$) is reflected off the surface of an SLM, and the plane of the SLM is imaged onto the back entrance of a 40mm focal length lens. The light focused on the sample generates Brillouin photons which are back scattered and collected by the lens to be coupled into a single mode fiber. The output of the fiber is spectrally dispersed by a double stage VIPA spectrometer, and a feedback loop is used to enhance the intensity of the acquired Brillouin spectrum.

Following the protocol first presented by Booth et al [41], our process of determining the optimized SLM phase is illustrated in figure 1(b). As a basis for phase aberrations we used the set of orthogonal Zernike polynomials (normalized to $2\pi$) which are well suited for circular boundary conditions. At every step of the iterative process, we sequentially projected on the SLM one Zernike polynomial and its inverse, and measured the intensity metric response of the spectrum for both conjugated Zernike polynomial patterns (denoted by $B, B^*$). The coefficient of each Zernike polynomial was determined by the relative weight of the intensity squared difference $\Delta B = B - B^*$. We measured the response difference of the first twenty-one Zernike polynomials excluding the vertical and horizontal tilt phases which represent merely a lateral shift of the focal point. To minimize the influence of random intensity fluctuations on the phase optimization protocol, we set a threshold on the final coefficient amplitude and after performing a weighted average of the calculated Zernike coefficients we projected the final corrected phase on the SLM. Considering the uniform phase projected on the SLM at the first step of the iterative process, the total number of iteration is $2N + 1$, where $N$ is the number of selected Zernike polynomials.

The experimental configuration of our AO-Brillouin system is presented in figure 1(c). We expanded a single frequency laser beam of wavelength 660 nm (LaserQuantum) and transmitted it through a linear polarizer to ensure a phase-only spatial modulation. We then reflected the beam off the surface of a spatial light modulator (LCOS-SLM, Hamamatsu X10468-01), and imaged the SLM plane using a 4-f imaging system (L1, L2, f=200mm) onto the back entrance of lens L3 (f=40mm) focusing the beam onto the sample of interest. We adopted a dual axis confocal configuration. The dual-axis configuration has reduced collection efficiency compared to epi-detection, but it eliminates noise generated by back reflections and yields higher axial resolution for a given numerical aperture [51]. We collected the scattered light and coupled it into a single mode fiber, serving as a confocal pinhole. Light was then dispersed by the double-stage VIPA spectrometer with a 15GHz free-spectral-range [23], and the obtained Brillouin spectrum was detected by an EMCCD camera (Andor Ixon 897).

## Results

To characterize the improvement of our AO-Brillouin system, we prepared a phantom sample featuring repeated layers of two different adhesive tapes (layer thickness ~70µm). An aberration was generated by spreading a thin layer of glue on the exterior surface of the sample; a chamber filled with water was then attached to the interior side (figure 2(a)). We performed the optimization process as described above and found the optimal Zernike coefficients on the water section of the sample (figure 2(b) bar graph). The dominant aberrations in this case are coma and astigmatism which are typically introduced by refractive index mismatches and thus expected in our case. The iterative process enhanced the intensity of the water signal by ~2.5-fold as shown in figure 2(c). AO correction also improves spatial resolution because of a sharper focal point at the measured location. To evaluate the resolution improvement, we performed an axial scan of the layered sample, and measured the Brillouin shift at each axial location before and after the AO correction. The optimal axial resolution of our system was previously characterized to be 47 µm at FWHM, sufficient to observe the layered structure. As the aberration is introduced, the resolution of the system degrades to approximately 80 µm without the AO phase correction and the layers of the sample cannot be distinguished (figure 2(d), blue dots). However, when AO correction is applied, the layers can be clearly resolved (figure 2(d), orange dots) and the resolution of the system is enhanced to 57 μm, approaching the optimal performances.

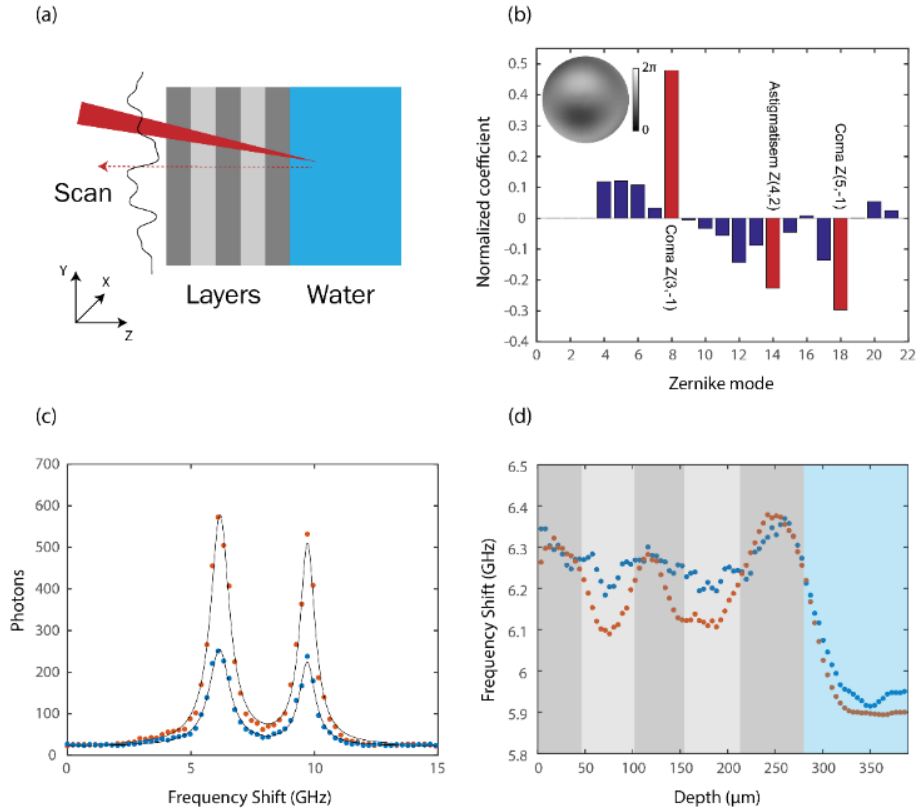

**Fig. 2.** Signal and resolution enhancement: (a) The phantom sample: a layered structure composed of two types of transparent adhesive tapes were placed in front of a home-built water chamber; an aberration was introduced by spreading a layer of glue on the sample. (b) The normalized Zernike coefficients obtained by the iterative algorithm optimizing the Brillouin signal of water through the fabricated glue aberration. The red bars represent the most significant modes (obtained by setting a threshold on the coefficients magnitudes), and the inset shows the phase map combining these modes. The iterative process was performed in ~40 $sec$. (c) The Brillouin spectrum showing the Stokes and anti-Stokes peaks of water without AO (blue dots) and with AO (orange dots), as well as a double Lorentzian fit (black line). (d) An axial scan of the sample with AO (orange dots) and without AO (blue dots).

In this measurement the optimization routine was performed only once throughout the entire axial scan. The region in which the optimized phase is still effective at correcting the aberration defines the isoplanatic patch, which is sample dependent. In the case of this phantom sample, the axial isoplanatic patch covered the entire sample and we observed 50% signal enhancement up to 300 microns away from the corrected location. A large isoplanatic axial range is expected in the case of a single layer transparent aberration; for comparison, within scattering biological tissues such as the mouse brain the isoplanatic correction volume has been previously found to be approximately 100 $\mu m^3$ [52, 53].

To demonstrate the advantage of our system within biological samples we performed an axial measurement through the cornea and aqueous humor of a fresh porcine eye (figure 3(a)). Although the cornea is transparent, the signal intensity drops considerably as a function of depth due to aberrations (figure 3(b) blue dots). In these conditions, the AO correction increased Brillouin signal intensity up to more than 2-fold (figure 3(b) orange dots). The enhancement became more significant at greater depths, as expected, given the increased signal degradation due to aberrations introduced by the cornea (figure 3(c)).

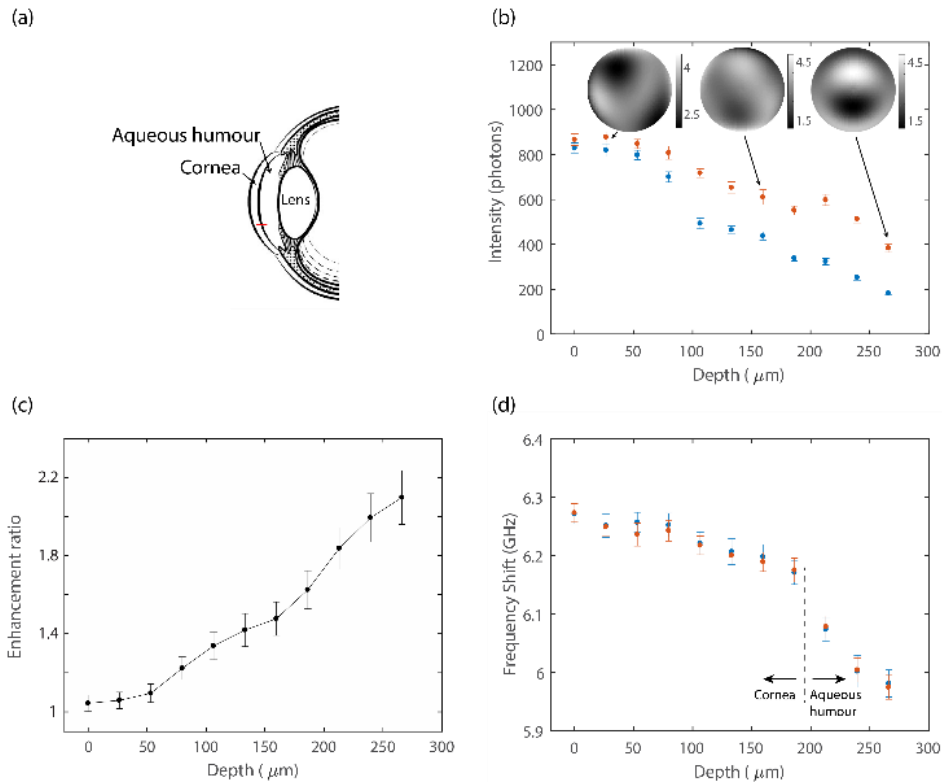

**Fig. 3.** Cornea and aqueous humor scan: (a) Structure of the porcine eye and the scanned region (red line). (b) The intensity of the Brillouin signal at various depths 27 microns apart throughout the cornea and aqueous humor of a fresh porcine eye. The optimization was performed at each location and three representative phase maps are shown. At higher depths the aberrations where considerably stronger and the signal was enhanced by more than 2-fold when AO was applied (orange dots) vs the uncorrected signal intensity (blue dots). (c) Signal enhancement as a function of measurement depth. (d) The Brillouin shift at every location in (b), with (orange dots) and without (blue dots) AO correction. Error bars represent the standard deviation of twenty measurements.

---

In figure 3(d) the Brillouin frequency shift is presented for every optimized location with (orange dots) and without (blue dots) AO correction. The difference in the Brillouin shift values falls within the standard deviation of the measurement. This result is relevant for practical purposes; an active element placed within the optical path can change location of the focal point while converging towards the optimal signal enhancement. Particularly, in Brillouin spectroscopy, the signal is not confined to a point object as in fluorescence microscopy but can originate from any point within the illuminated region (as in a so-called fluorescent sea). As a result, because the active element can easily shift the illumination focal point (e.g. our SLM can axially dislocate the focal point by hundreds of microns), when the signal intensity has a spatial gradient as is the case along the z-axis of the cornea sample, the tendency of the system to shift the measured location along the intensity gradient can be significant. In our experiment, the dislocation of the measured point was negligible, and the AO phase projection provided similar Brillouin shift values as the uncorrected scenario (figure 3(d)). We further confirmed this result by repeating the iterative process many times at a single location of the cornea and obtaining a variation in Brillouin shift of less than 10MHz, i.e. less than the single-point shift precision of our spectrometer. The reason we were able to avoid such measurement artifact is due to the confocal gating built within the setup, i.e. while the optimization process was performed on the illumination arm of the system, the collection path was kept fixed. Under these circumstances, a dislocation of the focal point will result in a mismatch between the illuminated point and the confocal pinhole which will decrease the signal intensity.

# Discussion

In the past years, a lot of progress has been made towards reducing noise elements of Brillouin spectral measurements so that shot-noise limited measurements are now available within scattering and aberrating biomaterials [19-23]. Nonetheless, Brillouin low signal intensity still imposes a restriction for many biological applications. To address this issue, improvements in the throughput of spectrometers [24] or stimulated scattering processes [26, 27] are being developed. Despite the promising applications enabled by these methods, the signal degradation due to optical aberrations is a fundamental problem that has remained unsolved. In this work we presented an AO-Brillouin confocal system, designed to enhance the signal and resolution of Brillouin-based elasticity mapping. Our method can be combined with any previously mentioned technique to optimize the detection efficiency of Brillouin photons.

AO has been extensively implemented in optical microscopy to provide optimal resolution and high SNR images. Here we obtained a 2.5-fold signal enhancement and a 1.4-fold resolution improvement. Although the enhancement factors are highly sample dependent, our enhancement results are consistent with previously reported AO studies for a variety of optical modalities applied to phantom samples as well as brain and ocular tissue [32, 50, 52]. Specifically, in Brillouin spectroscopy the signal can arise from any point in the illuminated region (as in a fluorescent sea) in which lower enhancements are expected [52].

Besides nicer looking images/spectra obtained by AO, the more important scenarios of AO applications are those where AO is essential to enable a measure by overcoming a fundamental barrier. As an example, we consider the case where the SNR of the acquired spectrum is less than one due to aberrations. In this scenario, the signal can't be recovered by increasing the integration time of the measurement or averaging many acquired spectra. To demonstrate this, we used a glass bottle filled with methanol as a test sample, and introduced an aberration on the glass by spreading a layer of glue (figure 4(a)). Because of the low SNR caused by aberrations and poor Brillouin gain of glass [54], measuring the Brillouin signal of glass through the aberration was impossible and averaging over many acquired spectra didn't increase the SNR of the measurement at all (figure 4(b), left panel). To demonstrate the power of AO correction in this scenario, we performed the optimization process on the methanol (which has high Brillouin gain) near the glass-methanol interface, i.e. within the isoplanatic patch. After obtaining the appropriate phase correction, we axially translated the sample to measure the Brillouin signature of glass and obtained a spectrum with SNR>1 that could be averaged for proper spectral analysis (figure 4(b), right panel). Under these conditions, AO offered a unique solution to enable spectral analysis as seen by the line plot in figure 4(c).

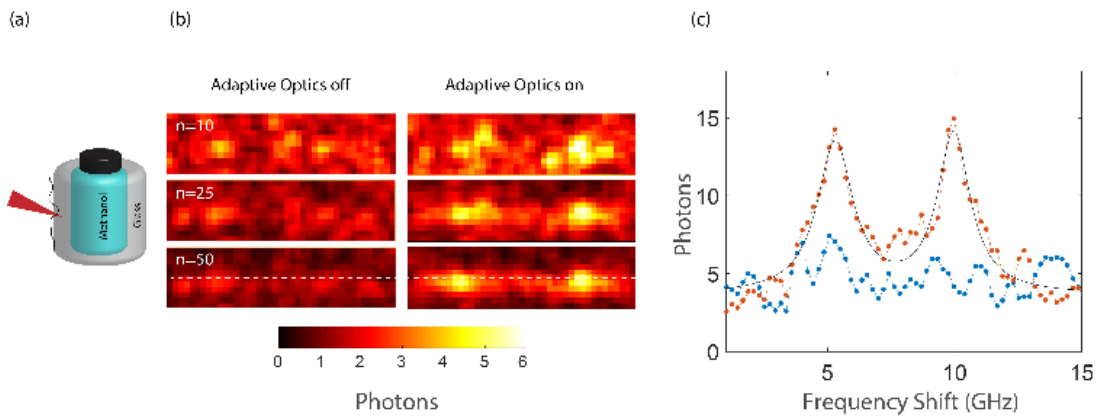

**Fig. 4.** AO-enabled measurements at low SNR: (a) Schematics of the experimental configuration. (b) Average of many acquired spectra of glass without AO (left) and with AO (right). A Gaussian filter (3x3 pixel, σ=2) was applied to the data. (c) Line plot of the area between the white dashed lines in (b). The spectrum of glass is not visible even when 50 frames are averaged if AO is not applied (blue); however, with AO correction on, the spectrum can be easily measured (orange).

A common consideration when applying AO to correct aberrations is whether correction is needed at both the illumination and detection paths. Confocal modalities can benefit from a double path correction [28, 30], and it was shown that a single active device is sufficient to correct simultaneously for both the illumination and collection paths [41]. Here we used the AO system to correct only for the illumination path for two main reasons. By keeping the collection path fixed during the iterative optimization process we ensured that the SLM doesn't shift the location of the measured point. In addition, while the finite reflectivity of the LCOS-SLM (~60 %) does not introduce net losses when placed in the illumination path because it can be compensated by increasing the illumination intensity, a second pass or a second device introduced in the collection path will lead to a significant signal loss which can't be retrieved.

As previously mentioned, in this work we adopted the indirect approach to AO correction and enhanced the Brillouin signal through an iterative process based on the acquired spectra. This approach is made possible by the rapid acquisition times characteristic of VIPA-based spectrometers which can be as low as 50 ms. Nevertheless, in the presence of aberrations, the indirect AO correction results in an overall optimization process of tens of seconds; this is clearly a drawback of this approach that makes this method suitable to non-absorbing samples where long illumination times are not a concern or to samples where isoplanatic regions are large enough that the iterative process needs to only be performed once for the entire scanned region [52, 53]. For other samples, a direct AO approach where aberrations are rapidly measured and corrected is required; to do so, a suitable 'guide-star' such as a fluorescent bead needs to be inserted into the sample at various locations, alternatively, scattering/reflections from internal structures within the samples can be used as 'guide-stars'. The performances of the direct and indirect AO implementations are expected to be similar, with the exception that indirect approaches can be applied to more scattering samples where the wave-front provided by the 'guide-star' has been shown to not be viable [30].

## Conclusion

In recent years, AO techniques have been extensively implemented in a variety of optical modalities showing promising results towards deep imaging within biological specimens. Here we demonstrated for the first time an AO-Brillouin integrated spectrometer, to correct for aberrations and enable high SNR spectral measurements with enhanced resolution. We tested our system using phantom samples as well as biological specimens such as the cornea of the eye, and presented an improvement of both signal intensity and resolution. As optical elastography modalities are widely being adopted for biomechanical studies, our system can enable elasticity mapping with higher precision and deeper within specimens thus potentially extending the applications of these techniques to more opaque samples.